\documentclass[letterpaper,twocolumn,
prl
,superscriptaddress,email
]{revtex4}

\usepackage{soul}
\usepackage[pdftex]{graphicx}
\usepackage{amssymb}
\usepackage{mathrsfs}
\usepackage{amsmath,amsfonts,latexsym}
\usepackage{array,tabularx,xcolor}
\usepackage{dcolumn} 
\usepackage[normalem]{ulem}
\usepackage{comment}
\usepackage{bm}

\usepackage{doi}
\usepackage{hyperref}

\hypersetup{
   colorlinks=true,
   linkcolor=magenta,
   citecolor=magenta,
   filecolor=magenta,
   urlcolor=red,
}

\newcommand{\redc}[1]{{\color{black}#1}}
\newcommand{\bluec}[1]{{\color{black}#1}}

\newcommand{\ket}[1]{| #1 \rangle}
\newcommand{\bra}[1]{\langle #1 |}
\newcommand{\tket}[1]{$|#1\rangle$}

\begin{document}

\title{A solid-state single-photon filter}

\author{L. de Santis}
\affiliation{Centre de Nanosciences et de Nanotechnologies, CNRS, Univ. Paris-Sud, Universit\'e Paris-Saclay, C2N -- Marcoussis, 91460 Marcoussis, France}

\author{C. Ant\'on}
\affiliation{Centre de Nanosciences et de Nanotechnologies, CNRS, Univ. Paris-Sud, Universit\'e Paris-Saclay, C2N -- Marcoussis, 91460 Marcoussis, France}

\author{B. Reznychenko}
\affiliation{CEA/CNRS/UJF joint team ``Nanophysics and Semiconductors", Institut N\'eel-CNRS, BP 166, 25 rue des Martyrs, 38042 Grenoble Cedex 9, France, Universit\'e Grenoble-Alpes \& CNRS, Institut N\'eel, Grenoble, 38000, France}

\author{N. Somaschi}
\affiliation{Centre de Nanosciences et de Nanotechnologies, CNRS, Univ. Paris-Sud, Universit\'e Paris-Saclay, C2N -- Marcoussis, 91460 Marcoussis, France}

\author{G. Coppola}
\affiliation{Centre de Nanosciences et de Nanotechnologies, CNRS, Univ. Paris-Sud, Universit\'e Paris-Saclay, C2N -- Marcoussis, 91460 Marcoussis, France}

\author{J. Senellart}
\affiliation{Systran -SA - Rue Feydeau - 75002 Paris, France}
\author{C. G\'omez}
\affiliation{Centre de Nanosciences et de Nanotechnologies, CNRS, Univ. Paris-Sud, Universit\'e Paris-Saclay, C2N -- Marcoussis, 91460 Marcoussis, France}

\author{A. Lema\^itre}
\affiliation{Centre de Nanosciences et de Nanotechnologies, CNRS, Univ. Paris-Sud, Universit\'e Paris-Saclay, C2N -- Marcoussis, 91460 Marcoussis, France}

\author{I. Sagnes}
\affiliation{Centre de Nanosciences et de Nanotechnologies, CNRS, Univ. Paris-Sud, Universit\'e Paris-Saclay, C2N -- Marcoussis, 91460 Marcoussis, France}

\author{A. G. White}
\affiliation{Centre for Engineered Quantum Systems, Centre for Quantum Computation and Communication Technology, School of Mathematics and Physics, University of Queensland, Brisbane, Queensland 4072, Australia}

\author{L. Lanco}
\affiliation{Centre de Nanosciences et de Nanotechnologies, CNRS, Univ. Paris-Sud, Universit\'e Paris-Saclay, C2N -- Marcoussis, 91460 Marcoussis, France}
\affiliation{Universit\'e Paris Diderot -- Paris 7, 75205 Paris CEDEX 13, France}

\author{A. Auffeves}
\affiliation{CEA/CNRS/UJF joint team ``Nanophysics and Semiconductors", Institut N\'eel-CNRS, BP 166, 25 rue des Martyrs, 38042 Grenoble Cedex 9, France, Universit\'e Grenoble-Alpes \& CNRS, Institut N\'eel, Grenoble, 38000, France}

\author{P. Senellart}
\email{pascale.senellart@lpn.cnrs.fr}
\affiliation{Centre de Nanosciences et de Nanotechnologies, CNRS, Univ. Paris-Sud, Universit\'e Paris-Saclay, C2N -- Marcoussis, 91460 Marcoussis, France}
\affiliation{Physics Department, Ecole Polytechnique, F-91128 Palaiseau, France}

\date{\today}

\begin{abstract}

A strong limitation of linear optical quantum computing \cite{KLM} is the probabilistic operation of two-quantum bit gates \cite{cnot} based on the coalescence of indistinguishable photons \cite{hom}. A route to deterministic \redc{operation} is to exploit the single-photon nonlinearity of an atomic transition. Through engineering of the atom-photon interaction,  phase \redc{shifters} \cite{Tiecke:2014aa,Reiserer:2014ab}, \redc{photon} filters \cite{Dayan1062,Rosenblum:2016aa} and photon-photon gates \cite{ritter2016} have been demonstrated with natural atoms. Proofs of concept have been reported with semiconductor quantum dots, yet limited by  inefficient atom-photon interfaces \redc{and  dephasing}  \cite{Reinhard:2012aa,waksgate,loo2012,PhysRevA.90.023846}. \redc{Here we report on a  highly  efficient single-photon filter based on a large optical non-linearity at the single photon level, in a near-optimal quantum-dot  cavity interface 
\cite{dousse2008,Somaschi2016}. When  probed with  coherent light  wavepackets, the device shows a record nonlinearity threshold around $0.3\pm0.1$ incident photons. We demonstrate that  directly reflected pulses consist of 80\% single-photon 
Fock state and that the two- and three-photon components are strongly suppressed  compared to the single-photon one.} 
\end{abstract}

\maketitle


Photons are the 
natural choice  
for 
connecting the---possibly distant---nodes of a quantum network  \cite{Kimble:2008aa}. The last decades have seen spectacular achievements using photonic quantum technologies, including teleportation \cite{teleportation}; linear computing of physical and chemical systems \cite{physical}, and quantum simulations of classically untractable problems such as \textsc{BosonSampling} \cite{boson1,boson2}.
Scaling quantum photonic technologies requires advances in three areas: bright and pure sources of indistinguishable and entangled single photons; high quantum-efficiency single-photon detectors; and efficient two-photon quantum gates. 
Impressive progress has been reported for both sources and detectors in the last few years, respectively using semiconductor quantum dots in cavities \cite{Somaschi2016,ding2016} and superconducting nanowire detectors \cite{swn}. 
However, quantum information protocols  are still using probabilistic techniques---based on the coalescence of indistinguishable photons \cite{KLM}---with typical success rates of 1/4 for a quantum relay \cite{relay} and  1/6--1/9 for a controlled-\textsc{not} gate \cite{cnot}. 

Such \bluec{linear schemes} cannot be scaled: recognising this, there have been many proposals for realising entangling-gates which achieve efficiency by operating in the nonlinear optical regime \cite{NLgate1,NLgate2}
. The paradigm for such nonlinear interactions is an optical transition in an atomic system: a first photon saturates the transition allowing the deterministic transmission of a second one \cite{PhysRevLett.75.4710,Chang:2014aa}.
\redc{A device providing a  deterministic photon-photon interaction  \cite{hu2008,bonato,dayanPRA} requires a perfect atom-photon interface and should be operated  with light wavepackets and without any  post-selection for use in quantum network applications.}
Coupling natural atoms to cavities \cite{Reiserer:2014aa,Tiecke:2014aa} or  waveguides \cite{Dayan1062} has recently allowed the demonstration of efficient atom-photon interfaces used to demonstrate photon filtering \cite{Rosenblum:2016aa, lukin2} and photon-photon deterministic gates \cite{ritter2016}. First explored using continuous-wave light fields \cite{Tiecke:2014aa}, recent works have 
reported gates and filters \redc{ operating on photon wavepackets}  \cite{Shomroni903,Rosenblum:2016aa}. 

Demonstrating such nonlinearity  at the single photon level in solid-state micron size devices  offers the potential of scalability and integration. Optical nonlinearity with a threshold at the  level of \redc{8} incident photons has been observed in micropillar cavities, \redc{yet limited by  dephasing} 
 \cite{loo2012}.  Photon blockade \cite{Reinhard:2012aa,Javadi:2015aa,PhysRevA.90.023846, toshiba} and optical gates \cite{reinhard2,waksgate,waks2016} based on a single quantum dot (QD) 
in  
cavities or slow-light waveguides have been reported. \redc{ However, these works operated either in the continuous-wave regime \cite{jelenaOld, toshiba,Javadi:2015aa} or with strong post selection \cite{Reinhard:2012aa,reinhard2,waksgate,waks2016,Javadi:2015aa,PhysRevA.90.023846} to compensate for the inefficient coupling  between the incident light and the device optical mode. In such cases, a  crossed-polarization 
post selection of the light was typically  implemented 
to remove the strong uncoupled incident light.}

\redc{Here we  investigate the light directly reflected from a single InGaAs QD inserted in a  micropillar cavity when sending temporally shaped coherent wavepackets. A record low
nonlinearity threshold is obtained  at \bluec{$0.3\pm0.1$}  photon per pulse  sent on the device, owing to an optimal QD-cavity coupling and a marginal pure dephasing.}
\redc{The device  is a very efficient single-photon Fock-state 
filter  converting a coherent pulse into a highly non-classical wavepacket
without any post-selection}.

\begin{figure}[t]
\setlength{\abovecaptionskip}{-5pt}
\setlength{\belowcaptionskip}{-2pt}
\begin{center}
\includegraphics[width=1\linewidth,angle=0]{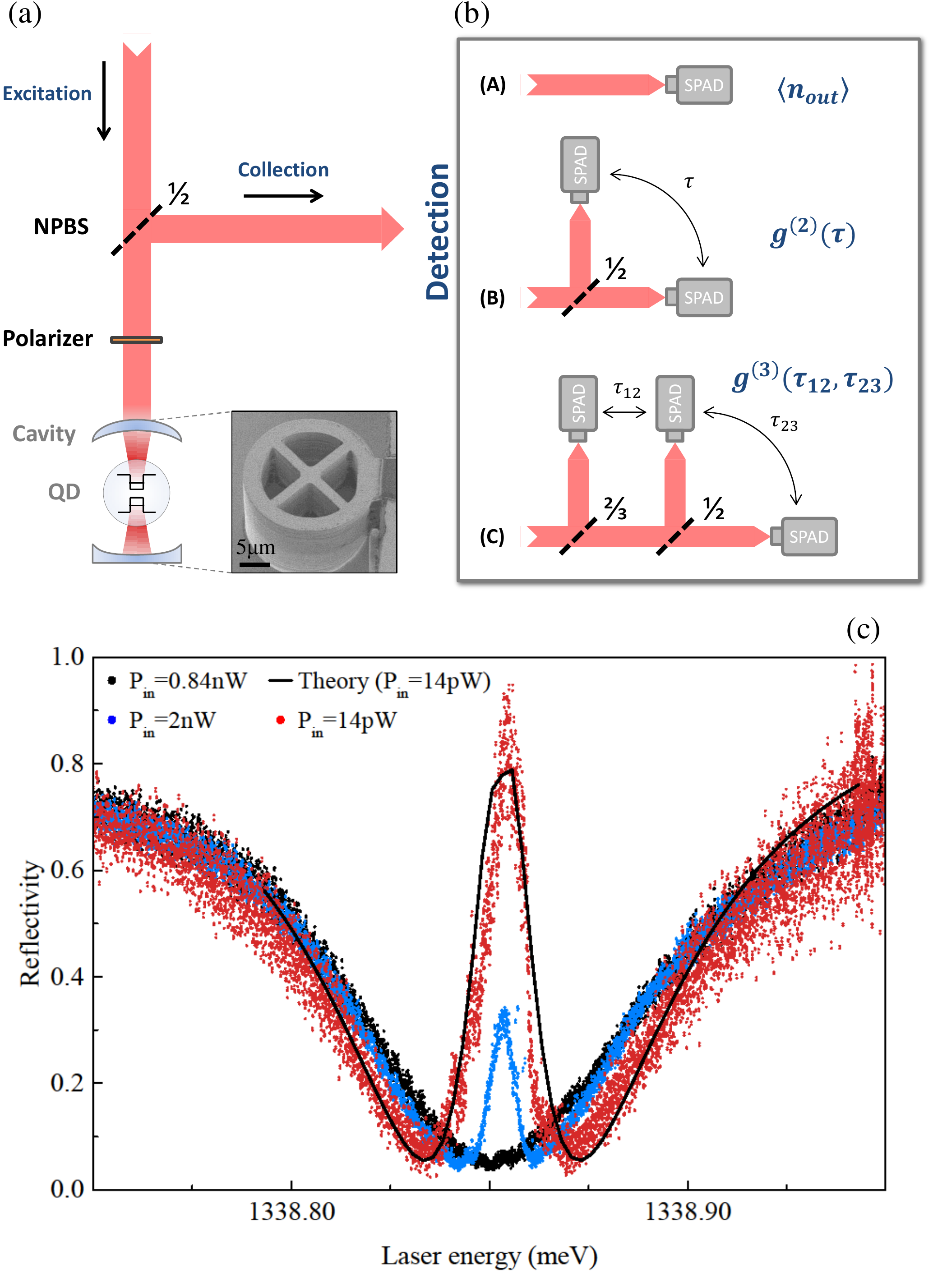}
\end{center}
\caption{{\bf  Experimental design and Characteristics of the device} (a) Experimental schematic: coherent photon wavepackets are sent on a quantum-dot cavity device. The insert shows a scanning electron microscopy image of the device. (b) The light directly reflected is analyzed using one of the three configurations labelled A, B and C on the detection line.  (c) Reflectivity spectrum of device obtained under continuous wave excitation for three excitation conditions. Symbols: red (14 pW), blue (840 pW), black (2 nW); Black line: theoretical adjustment for the lowest excitation power. }
\label{fig:fig1}
\end{figure}

In our device the cavity is centered on the QD with 50 nm accuracy using the in-situ optical lithography technique \cite{dousse2008,Nowak:2014aa} and connected through four ridges to a large frame where metallic contacts are defined. The contact allows fine tuning of the QD-cavity spectral resonance through the confined Stark effect by application of a bias \cite{Nowak:2014aa,Somaschi2016}.
The QD transition under study is a neutral exciton state which corresponds to two linearly polarized dipoles slightly split in energy by the  fine structure splitting $\Delta_\mathrm{FSS}$. 
The device 
is kept at 9 K in a closed-circuit helium cryostat, and excited by a linearly polarized tunable continuous-wave or pulsed  laser focused by a microscope objective. 
We use the objective to collect the signal directly reflected from the sample in the \emph{same} optical mode as the incident wave packets, i.e. same spatial mode and polarization, Fig. \ref{fig:fig1}(a). This output signal is coupled to a single-mode optical fiber (SMF) and coupled to one, two, or three fiber-coupled single-photon avalanche diodes (SPADs) to respectively measure the reflectivity, the second-order, or the third -order intensity-correlation functions (Fig. \ref{fig:fig1}(b)). 

The figures of merit of the QD-cavity device are measured through  reflectivity measurements using a tunable continuous-wave laser (1 MHz linewidth) \cite{loo2012}. When the QD is detuned from the cavity mode, the reflectivity shows a cavity dip that renders information on the cavity mode polarization, linewidth and output coupling efficiency. The cavity modes are linearly polarized along two directions  labelled $H$ and $V$ with a total cavity damping rate $\kappa{=}90$ $\mu$eV 
(quality factor of 14000) and a mode splitting of 70 $\mu$eV. The ratio between the top-mirror damping rate and the total damping rate $\kappa$, corresponding to the probability for cavity photons to escape the cavity through the top port (output coupling efficiency), is $\eta_{\textrm{top}}{=}64\%$. 
Considering the asymmetric design of Bragg mirrors (30/20 for bottom and top mirrors), we deduce that the remaining escape channels for the photons outside the cavity are 10\% through the bottom mirror and 26\% through the lateral ridges. An input coupling efficiency above 95\% is also extracted by measuring the overlap between the incident field and the mode profile.

When exciting along \redc{$H$} and bringing the QD in resonance with the cavity mode  at low excitation power ($P_{\textrm{in}}{=}14$ pW), 
a strong signal coming from the coherent response of the QD is observed at the center of the cavity dip, see Fig. \ref{fig:fig1}(c), 
with reflectivity as large as 90\%.
The evolution of the system is computed with a master equation involving the three level states of the exciton and both cavity modes and the different lossy modes (
see methods).  

A theoretical adjustment to the reflectivity spectrum gives the coupling constant between the cavity mode and the exciton transition $g{=}19$ $\mu$eV \ 
and a radiatively limited total exciton dephasing rate $\gamma{=}\frac{\gamma_{\textrm{sp}}}{2}{+}\gamma^* {=} 0.3 {\pm} 0.05 \mu$eV 
with the  spontaneous emission rate $\gamma_{\textrm{sp}}{=}0.6{\pm}0.1 \mu$eV 
and a negligible pure dephasing rate $\gamma^*{=}0.03{\pm}0.03\ \mu$eV. 
From this spectrum, one can also deduce the  exciton fine structure splitting $\Delta_{\mathrm{FSS}}{=}3 \mu$eV 
and the relative orientation of the QD and cavity axes $\theta{=}15{\pm}5^\circ$. 
The device operates in the weak coupling regime with a state of the art cooperativity $C{=}g^2/(\kappa\gamma) {\approx} 12$ 
corresponding to a Purcell factor of $F_P{=}2C{=}24$. 
\redc{Such large cooperativity ensures that the QD  exciton radiative\bluec{ly decays} into the cavity mode with a probability  $\beta{=}0.97$.} 
The very high value of $\beta$, input coupling efficiency of 95\% and output coupling efficiency 64\% shows that the device under study is close to the one dimensional atom situation as recently evidenced \cite{GieszNatCom} and reported here for another device. 
When increasing the excitation power, a continuous saturation of the exciton transitions is observed as shown in Fig. \ref{fig:fig1}(c). It corresponds  to a record low critical intracavity photon number $n_c{=}\frac{\gamma_{\textrm{sp}}^2}{8g^2}{\approx} 10^{-4}$ \redc{for the onset \bluec{of} the non-linearity in a continuous wave measurement}. 

\begin{figure}
\setlength{\abovecaptionskip}{-5pt}
\setlength{\belowcaptionskip}{-2pt}
\begin{center}
\includegraphics[width=1\linewidth,angle=0]{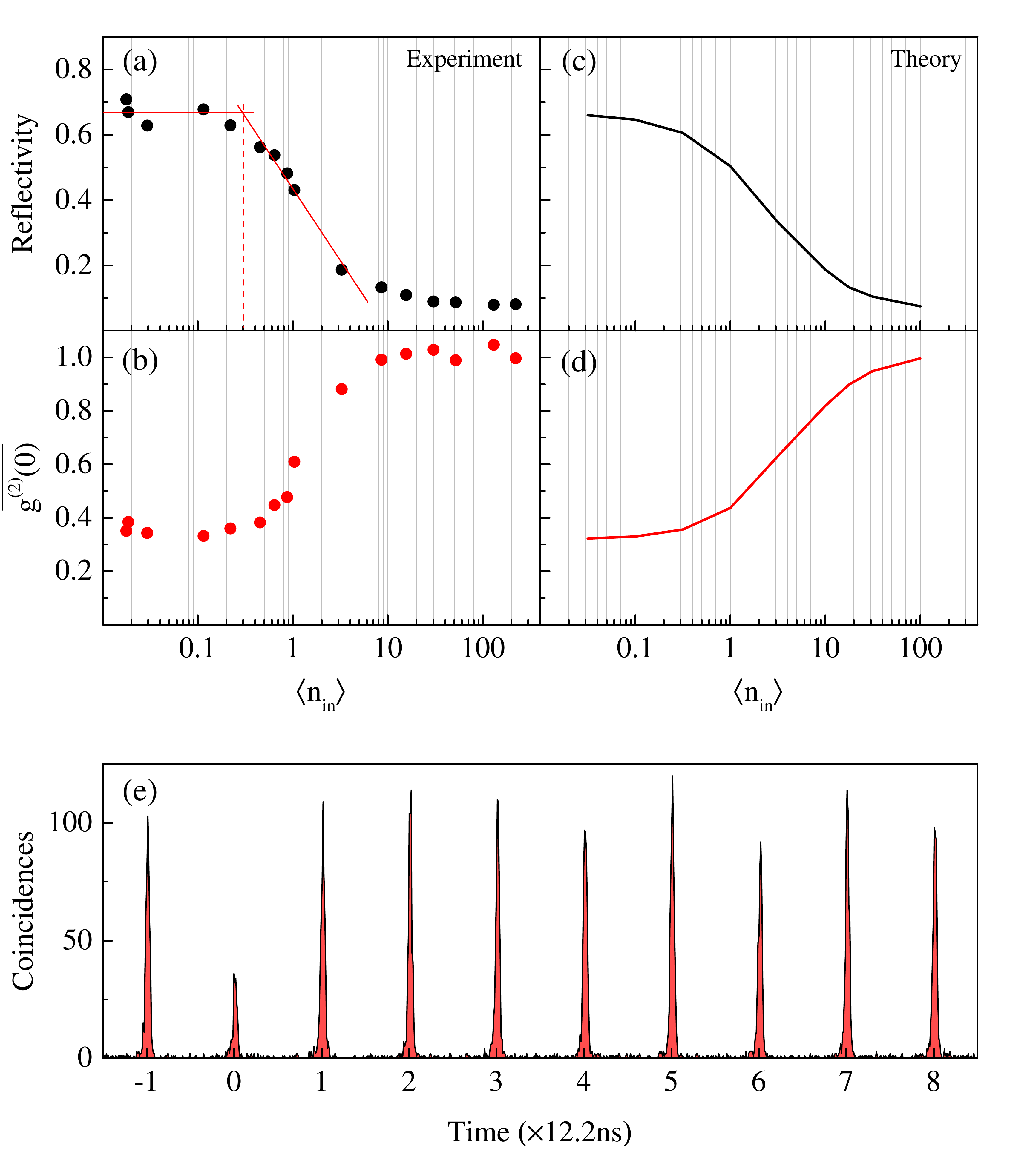}
\end{center}
\caption{{\bf Single photon nonlinearity and Second-order correlation measurements. }(a) Measured and (c) calculated reflectivity of a $125\ ps$ pulse (resonant at the QD transition) as function of the incident average photon-number $\langle n_{\textrm{in}}\rangle$. The straight lines in panel (a) are guide to the eyes showing the nonlinear threshold. (b) Measured and (d) calculated time-integrated second-order correlation function, $\overline{g^{(2)}(0)}$, as a function of $\langle n_{\textrm{in}}\rangle$. (e) Measured two-photon coincidences for $\langle n_{\textrm{in}}\rangle=0.1$  }
\label{fig:fig2}
\end{figure}

We now operate the device in the regime most suitable for quantum networks, directly manipulating light wave-packets.
From this point on, we probe the cavity with coherent light pulses of controlled temporal profile and average photon number per pulse $\langle n_{\textrm{in}}\rangle$.
We spectrally shape the 3 ps light pulses from a Titanium-Sapphire laser (82 MHz repetition rate) to match the exciton transition energy and radiative lifetime (125 ps).
The average photon number is deduced from the incident power sent on the device $P$ through $\langle n_{\textrm{in}}\rangle {=}\frac{P}{\Gamma_{rep} \hbar \omega_{laser}}$ 
where $\Gamma_{rep}$ and \bluec{$\omega_{laser} $ are the laser repetition rate and frequency}. Fig. \ref{fig:fig2}(a) shows the reflected intensity---measured using the setup sketched in Figure 1(b), Configuration A---normalized to the incident one as function of $\langle n_{\textrm{in}}\rangle$.
In the low average photon number regime, the coherent response of the exciton dominates and a very high reflectivity of $R_{max}=68\pm 2\%$ is observed, a value slightly reduced as compared to the continuous wave measurement (figure 1(c)) due to the finite  spectral width of the pulses.   At high excitation power,  the reflectivity saturates around $R_{min}=8 {\pm} 1\%$. 
The observation of \bluec{such a} large contrast  for the nonlinearity ---\redc{as large as $\frac{R_{max}-R_{min}}{R_{min}}\approx 8.5$--- represents a strong improvement to previous solid-state implementations where best contrasts were around 1.1} \cite{Javadi:2015aa,loo2012}. 
Similarly, the onset of the QD nonlinearity is observed around $\langle n_{\textrm{in}}\rangle = 0.3\pm0.1$, a value 25-40 times smaller than the previous state of the art \cite{loo2012}.
\redc{Such high contrast and nonlinearity at the very single-photon limit guarantee both an efficient reflexion of the single-photon component  and efficient suppression of the higher photon numbers, central features to build deterministic gates and single-photon filters. }

The second order correlation function $g^{(2)}(\tau)$ of the signal is measured through correlation measurements at the output of a 50/50 fibered beam splitter, see Figure 1.b., Configuration B. 
The \redc{time-integrated} normalized area of the zero delay peak \bluec{denoted $\overline{g^{(2)}(0)}$,} is plotted  in Fig. \ref{fig:fig2}(b) as function of  $\langle n_{\textrm{in}}\rangle$. For  $\langle n_{\textrm{in}}\rangle {\lesssim} 0.1$, 
the reflected signal is strongly anti-bunched \redc{as shown in figure 2.e.}, with a $\overline{g^{(2)}(0)} {\approx} 0.35$ 
showing that the  reflected beam is dominated by single-photons. The present results contrast with former studies of solid-state photon blockade where---despite a laser suppression scheme based on  cross polarization---limited antibunching in the $\overline{g^{(2)}(0)} {=} 0.6-0.9$ range were reported \cite{jelenaOld,reinhard2,Javadi:2015aa}. 
When increasing the incident photon number, the reflected light beam statistics progressively evolves from sub-poissonian to poissonian  ($\overline{g^{(2)}(0)}{=}1$): 
for $\langle n_{\textrm{in}}{\rangle}>10$, 
where the QD transition is saturated, the reflected field is dominated by the coherent component. 

\begin{figure}
\setlength{\abovecaptionskip}{-5pt}
\setlength{\belowcaptionskip}{-2pt}
\begin{center}
\includegraphics[width=1\linewidth,angle=0]{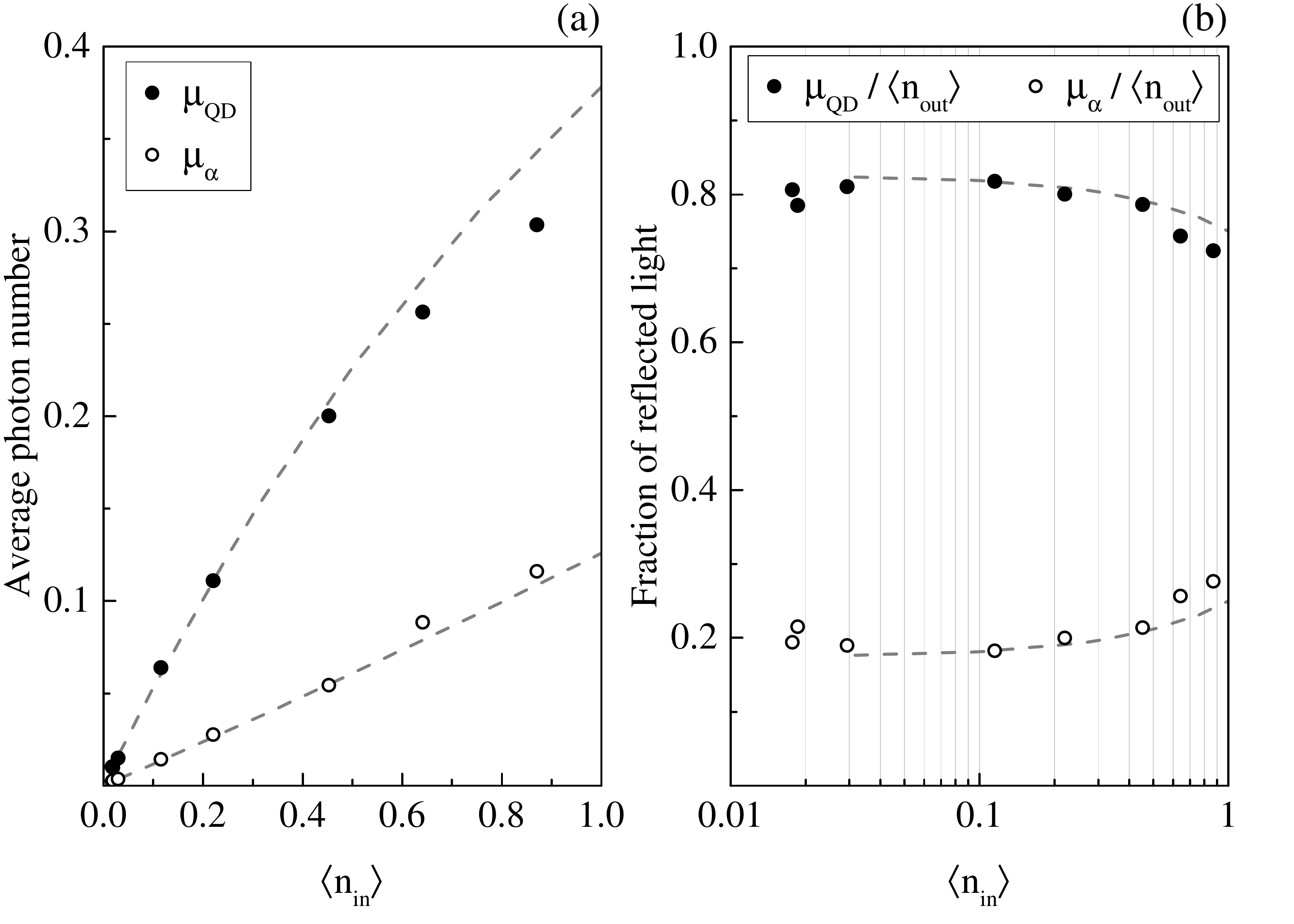}
\end{center}
\caption{{\bf Single-photon filtering. }(a) Average photon number vs $ \langle n_{\textrm{in}}\rangle$ (linear scale) for ($\bullet$) the single Fock-state component , $\mu_{QD}$, and ($\circ$) the coherent component, $\mu_{\alpha}$, of the output field. (b) Fraction of reflected light vs $ \langle n_{\textrm{in}}\rangle$ (log. scale) for ($\bullet$) single-photon $\frac{\mu_{QD}}{\langle n_{\textrm{out}}\rangle}$, and ($\circ$) coherent light, $\frac{\mu_{\alpha}}{\langle n_{\textrm{out}}\rangle}$. The dashed lines present the theoretical values deduced from the calculation presented in figure \ref{fig:fig2}(c-d).}
\label{fig:fig3}
\end{figure}

\begin{figure*}
\setlength{\abovecaptionskip}{-5pt}
\setlength{\belowcaptionskip}{-2pt}
\begin{center}
\includegraphics[width=0.9\linewidth,angle=0]{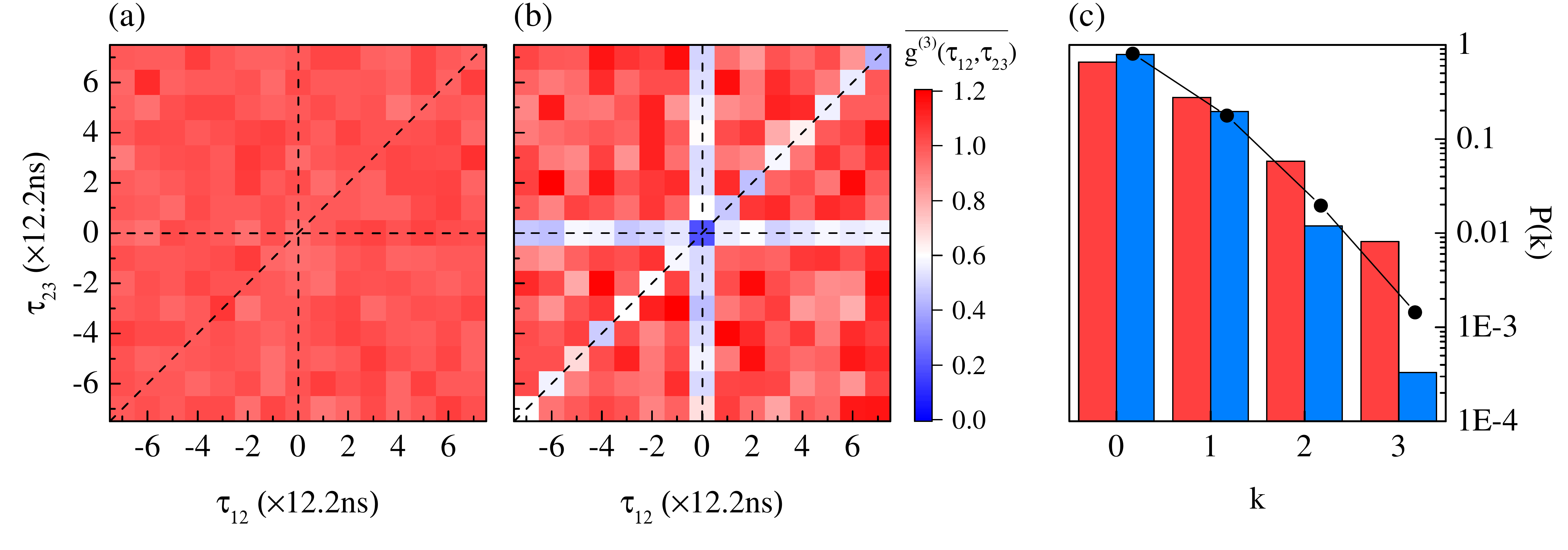}
\end{center}
\caption{{\bf Multi-photon state suppression. }(a) Correlation map of the incident field $\overline{g^{(3)}(\tau_{12},\tau_{23})}$. (b) Correlation map of the reflected field $\overline{g^{(3)}(\tau_{12},\tau_{23})}$ for $\langle n_{\textrm{in}}\rangle\approx0.6\pm0.1$. The dashed lines correspond to $\tau_{12}=0$, $\tau_{23}=0$ and $\tau_{12}=\tau_{23}$. (c) Occupation probabilities for the incident field $P_{\textrm{in}}(k)$ (red) and the output field  $P_{\textrm{out}}(k)$ (blue). The black symbols shows the Poisson distribution corresponding to the average output photon number $\langle n_{\textrm{out}}\rangle$.}
\label{fig:fig4}
\end{figure*}

Figure \ref{fig:fig2}(c) and (d) present the calculated reflectivity and $\overline{g^{(2)}(0)}$ using the parameters extracted from the continuous wave reflectivity measurements. Gaussian temporal profiles are used for the incident pulses. A very good overall agreement is obtained with the experimental observations. 
Note that the contrast and nonlinearity threshold in both reflectivity and  $\overline{g^{(2)}(0)}$ measurements depend significantly on the pulse temporal length as shown in supplementary figure \redc{S1 for another device.}

We now analyze the operation of our device as a single-photon Fock-state filter. 
At low incident photon number, the output field  is a mixture of light directly reflected from the cavity - presenting a Poisson statistic - and  light re-emitted by the QD - made of vacuum and single photons. To deduce the fraction of each contribution from our correlation measurements, we use the  probability-generating function formalism \cite{PhysRevA.88.013822} \redc{where the evaluation of the second or third derivative of a generating function $G_{total}(s)$, calculated at $s{=}1$, gives the second or third order intensity correlation function of the signal} \cite{PhysRevA.88.013822}. \bluec{The total photon number in the output field $\langle n_{\textrm{out}}\rangle$ presents two contributions $\langle n_{\textrm{out}}\rangle=\mu_{QD}+\mu_{\alpha}$, where $\mu_{QD}$ is the average photon number coming from the QD and $\mu_{\alpha}$ the average photon number for the coherent component. The probability-generating function for the light re-emitted by the QD is $G_{QD}(s) {=} (1{-}\mu_{QD}){+}s\mu_{QD}$ , and $G_{\alpha}(s) {=} e^{-\mu_{\alpha}(1{-}s)}$ for the coherent part}. The generating function for the total reflected mixture is given by $G_{total}(s) {=} G_{QD}(s) G_{\alpha}(s)$ 
and its second derivative $G''_{total}(s)$ gives $\overline{g^{(2)}(0)} {=} \frac{G''_{total}(s)|_{s{=}1}}{\langle n_{\textrm{out}}\rangle^2}$ 
\cite{PhysRevA.88.013822}. From the total reflected count rate, directly linked to the reflected total average photon number $\langle n_{\textrm{out}}\rangle{=}\mu_{QD}{+}\mu_{\alpha}$ 
and the measured $\overline{g^{(2)}(0)}$, we extract $\mu_{QD}$ and $\mu_{\alpha}$. 

Figure  \ref{fig:fig3}(a) presents $\mu_{\alpha}$ (open points) and $\mu_{QD}$ (filled points) as function of $\langle n_{\textrm{in}}\rangle$ as well as the theoretical values (dashed lines) deduced from the calculation presented in Fig. \ref{fig:fig2}(c-d). As expected, the light directly reflected by the cavity $\mu_\alpha$ linearly depends on the excitation power. In contrast, the QD response saturates when approaching $\langle n_{\textrm{in}}\rangle {\approx} 1$. 
For all values of $\langle n_{\textrm{in}}\rangle$, the reflected field is dominated by the QD field. Figure \ref{fig:fig3}(b) presents the fraction of single-photon $\frac{\mu_{QD}}{\langle n_{\textrm{out}}\rangle}$ and coherent light $\frac{\mu_{\alpha}}{\langle n_{\textrm{out}}\rangle}$ in the reflected beam and corresponding theoretical values. 
When sending a coherent beam on the device, the reflected field is shown to be 80\% single-photon. These observations, measured on the light directly reflected from the device without any additional polarization or temporal filtering, demonstrate that the device operates as a single-photon Fock state filter.

To get a better insight into the nature of reflected field, third-order intensity-correlation measurements are performed by splitting the signal to three detectors using two cascaded beam splitters, see Fig 1b, Configuration C.
The three-photon coincidences between detectors are recorded as a function of the time delay $\tau_{12}$ and $\tau_{23}$ between detectors 1 and 2  and detectors 2 and 3. The measurement provides a two-dimensional histogram of coincidences with peaks temporally separated by the repetition rate of the laser ($12.2$ ns). \bluec{The correlation map of $\overline{g^{(3)}(\tau_{12},\tau_{23})}$ is displayed in Fig. 4a-b: for each pixel of this map $\overline{g^{(3)}(\tau_{12},\tau_{23})}$ is} obtained by integrating the coincidence peaks over a temporal area $5 {\times} 5$ ns$^2$, and normalized to the average area of uncorrelated peaks (located at delays where $\tau_{12} {\neq} 0$, $\tau_{23} {\neq} 0$ and $\tau_{12} {\neq} \tau_{23}$). 
The dashed lines in Figure 4(a,b) 
\bluec{correspond to  $\tau_{12} {=} 0$, $\tau_{23} {=} 0$ or $\tau_{12} {=} \tau_{23}$, i.e. to 
a zero time delay between at least two detectors}. 

Figure 4(a) presents the correlation map for the incident field: a uniform pattern corresponding to $\overline{g^{(3)}(\tau_{12},\tau_{23})} {\approx} 1$ 
is observed confirming the Poisson statistic of the incident beam. Figure 4b presents the measurement obtained on the reflected beam \bluec{at $\langle n_{\textrm{in}}\rangle {\approx} 0.6\pm0.1$. A 
clear anti-bunching of $\overline{g^{(3)}(\tau_{12},\tau_{23})}$ is observed on the lines $\tau_{12} {=} 0$, $\tau_{23} {=} 0$ and $\tau_{12} {=} \tau_{23}$, for which $\overline{g^{(3)}(\tau_{12},\tau_{23})}\approx0.55$: this corresponds to a zero time delay between only two detectors , and thus is equivalent to a direct measurement of  $\overline{g^{(2)}(0)}\approx0.55$ \cite{PhysRevA.90.023846}. The pixel at the center of the map, however, corresponds to a zero time delay between all three detectors: this provides the value $\overline{g^{(3)}(0,0)} {=} 0.18$, showing a strong suppression of the 3-photon component of the incident field. }

From this measurement,  the occupation probabilities $P_{\textrm{out}}(k)$ of the $k$-photon state in the output field are deduced for $k\le3$ using four equations:\\
(i) the normalization of the probabilities ($1{=} \sum\limits_{k{=}0}^3 P_{\textrm{out}}(k)$),\\
(ii) the output mean average photon number ($\langle n_{\textrm{out}}\rangle{=}\sum\limits_{k=0}^3 k P(k)$),\\
(iii) the second order correlation function $g^{(2)}(0){=}\frac{(2P_{\textrm{out}}(2){+}6P_{\textrm{out}}(3))}{\langle n_{\textrm{out}}\rangle^2}$ and\\
(iv)  the third  order correlation function $g^{(3)}(0){=}\frac{6P_{\textrm{out}}(3)}{\langle n_{\textrm{out}}\rangle^3}$,\\
the last two expressions being valid since $P_{\textrm{out}}(k\ge 4) {\ll} P_{\textrm{out}}(k<4)$ \cite{Stevens:14}.
Figure 4c 
presents the occupation probabilities for the incident field $P_{\textrm{in}}(k)$ (red) and the output field  $P_{\textrm{out}}(k)$ (blue). The $\bullet$ 
symbols show 
the Poisson distribution corresponding to the average output photon number per pulse $\langle n_{\textrm{out}}\rangle$ and highlights 
the strong deviation from this statistics in the output field. While $P_{\textrm{out}}(1)/P_{\textrm{in}}(1) {\approx} 2/3$, 
\bluec{corresponding to a slight decrease of the 1-photon component}, a strong suppression ratio of the 2 and 3 photon components is observed with $P_{\textrm{out}}(2)/P_{\textrm{in}}(2) {\approx} 1/5$ and $P_{\textrm{out}}(3)/P_{\textrm{in}}(3) {=} 1/27$
: our device performs excellently as a multi-photon state suppressor. 

The photon-sorter device reported here provides a nonlinearity threshold and photon-sorting 
capability at the level of the best experimental realizations with single natural atoms coupled to optical waveguides or cavities \cite{Tiecke:2014aa,Reiserer:2014aa,Dayan1062,Rosenblum:2016aa}. Our fully integrated approach present the advantage of  not suffering from a limited trapping time \cite{Tiecke:2014aa,Reiserer:2014aa} or interaction time \cite{Dayan1062,Rosenblum:2016aa} of the atom  with the optical mode. In the present device geometry, the $k {\ge} 2$ 
photon components 
are transmitted through the four waveguides connected to the micropillar. To implement a two-photon 
gate, the cavity design could be engineered to ensure that 50\% of the light is coupled into a single in plane waveguide. Moreover, 
a single electron or hole spin  could be introduced in  controlled way in order to obtain \bluec{an} additional time scale---that of  the spin coherence time---needed to ensure the deterministic operation of the gate \cite{dayanPRA,hu2008,bonato}. 
Our results show that deterministic optical gates could be realized in micron-sized solid-state devices 
in the near future, a very promising perspective for boosting quantum photonic technologies that are today limited by probabilistic gates. 


\acknowledgments
\noindent {\bf Acknowledgments}
This work was partially supported by the ERC Starting Grant No. 277885 QD-CQED, the French Agence Nationale pour la Recherche (grant ANR SPIQE and USSEPP) the French RENATECH network,  a public grant overseen by the French National Research Agency (ANR) as part of the "Investissements d'Avenir" program (Labex NanoSaclay, reference: ANR-10-LABX-0035), 
the ARC Centres for Engineered Quantum Systems (Grant CE110001013), and Quantum Computation and Communication Technology (Grant CE110001027), and the Asian Office of Aerospace Research and Development (Grant FA2386-13-1-4070). AGW acknowledges support from a UQ Vice-Chancellor's Research and Teaching Fellowship. 

\noindent {\bf Authors contributions}
The experiments were conducted by L. d. S with help from N. S., C. A. and G. C. and suggestions from A.G. W. Data analysis was done by L. d. S., C. A. and J. S. The cavity devices were fabricated by N. S. from samples grown by A. L. and C. G. Etching was done by I.S. Theory was developed by B. R. under the supervision of A.A. with help from L. L. All authors participated to scientific discussions and manuscript preparation. This project was supervised by L.L., A.A. and  P. S.

\bibliographystyle{naturemag}

\pagebreak
\widetext
\begin{center}
\textbf{\large Methods Summary}
\end{center}
\setcounter{equation}{0}
\setcounter{figure}{0}
\setcounter{table}{0}
\setcounter{page}{1}
\makeatletter
\renewcommand{\theequation}{S\arabic{equation}}
\renewcommand{\thefigure}{S\arabic{figure}}
\renewcommand{\bibnumfmt}[1]{[S#1]}
\noindent {\bf Sample fabrication}

\noindent The  sample was grown by molecular beam epitaxy on a GaAs substrate. The light is confined in a semiconductor microcavity comprising a $\lambda$-GaAs spacer sandwiched between two asymmetric distributed Bragg reflectors. These mirrors are formed by alternating $\lambda/4$-thick layers of GaAs and Al$_{0.95}$Ga$_{0.05}$As, the top (bottom) mirror contains 20 (30) pairs enhancing the photonic losses from the top of the device. A dilute InGaAs self-assembled 
QD layer is inserted at the center of the cavity. The top and bottom mirror are postively and negatively doped to define a diode structure used to tune the QD energy.  After the epitaxial growth, the deterministic positioning and spectral matching of QD-cavity system has been realized through the in-situ lithography technique \cite{dousse2008,Nowak:2014aa} and dry pillar etching. The micropillar is connected to an outer circular shell by four ridges connected to metallic contacts, this allows 
electrical tuning of the QD transition \cite{Nowak:2014aa,Somaschi2016}.\\

\noindent {\bf Photon correlation measurements}

\noindent  Photon correlation experiments have been performed with a HydraHarp 400 autocorrelator working on its Time-Tagged Time-Resolved mode, using a pulsed Ti:Sapphire laser as synchronization signal. For  $\overline{g^{(3)}(\tau_{12},\tau_{23})}$ measurements, the output emission, collected in a SM fiber, has been connected to 
two cascaded fiber BSs (33:66 and 50:50 splitting-ratios, respectively, providing a balanced splitting-ratio in the three output ports connected to the fibered single-photon avalanche photodetectors. 
An example of three-photon coincidence raw histograms is shown in the supplementary figure \redc{S2}. \\


\noindent {\bf Theory}

\noindent The QD is modeled as a three-level system, both equally coupled to two orthogonal, quasi-resonant modes of the cavity. The fine structure splitting and the angle between QD and cavity natural axes are taken into account, as well as the finite spatial overlap between the cavity and the driving field, the mirrors finite transmittances and losses, and the finite detection efficiency which are modeled as additional terms in the Lindbladian/input-output equations. The driving light is modeled by a classical, time-dependent Hamiltonian (See Supplemental).
Introducing the output field $b_{out}$, the value of $g^{(2)}(0)$ was computed using the following formula:
\begin{align}\label{g2}
\overline{g^{(2)}(0)}=\frac{\int^\infty_{-\infty} dt \int^\infty_{-\infty} d\tau G^{(2)}(t,t+\tau)}{[\int^\infty_{-\infty} dt \langle b_{out}^\dag(t) b_{out}(t)\rangle]^2}
\end{align}
where 
\begin{align}
G^{(2)}(t_1,t_2)=\langle  \hat{b}_{out}^\dag(t_1) \hat{b}_{out}^\dag(t_2) \hat{b}_{out}(t_2) \hat{b}_{out}(t_1)\rangle,
\end{align}
where the two-time correlations functions are derived using the Quantum Regression Theorem.




\pagebreak
\widetext
\begin{center}
\textbf{\large Supplemental Materials: A solid-state single-photon filter}\\
\medskip
L. de Santis,$^1$ C. Ant\'on,$^1$ B. Reznychenko,$^2$ N. Somaschi,$^1$ G. Coppola,$^1$ J. Senellart,$^3$ C. G\'omez,$^1$ A. Lema\^itre,$^1$ I. Sagnes,$^1$ A. G. White,$^4$ L. Lanco,$^{1,5}$ A. Auffeves,$^2$ and P. Senellart$^{1,6}$
\medskip

\textit{$^1$ Centre de Nanosciences et de Nanotechnologies, CNRS, Univ. Paris-Sud, Universit\'e Paris-Saclay, C2N -- Marcoussis, 91460 Marcoussis, France\\
$^2$ CEA/CNRS/UJF joint team ``Nanophysics and Semiconductors", Institut N\'eel-CNRS, BP 166, 25 rue des Martyrs, 38042 Grenoble Cedex 9, France, Universit\'e Grenoble-Alpes \& CNRS, Institut N\'eel, Grenoble, 38000, France\\
$^3$ Systran -SA - Rue Feydeau - 75002 Paris, France\\
$^4$ Centre for Engineered Quantum Systems, Centre for Quantum Computation and Communication Technology, School of Mathematics and Physics, University of Queensland, Brisbane, Queensland 4072, Australia\\
$^5$ Universit\'e Paris Diderot -- Paris 7, 75205 Paris CEDEX 13, France\\
$^6$ Physics Department, Ecole Polytechnique, F-91128 Palaiseau, France\\}
\end{center}
\setcounter{equation}{0}
\setcounter{figure}{0}
\setcounter{table}{0}
\setcounter{page}{1}
\makeatletter
\renewcommand{\theequation}{S\arabic{equation}}
\renewcommand{\thefigure}{S\arabic{figure}}
\renewcommand{\bibnumfmt}[1]{[S#1]}
\renewcommand{\citenumfont}[1]{S#1}

\medskip
\medskip
\medskip
\medskip

\noindent {\bf Influence of the pulse length on the reflectivity contrast and second order correlation function. }
\medskip

\begin{figure}[h]
\setlength{\abovecaptionskip}{-5pt}
\setlength{\belowcaptionskip}{-2pt}
\begin{center}
\includegraphics[width=.80\linewidth,angle=0]{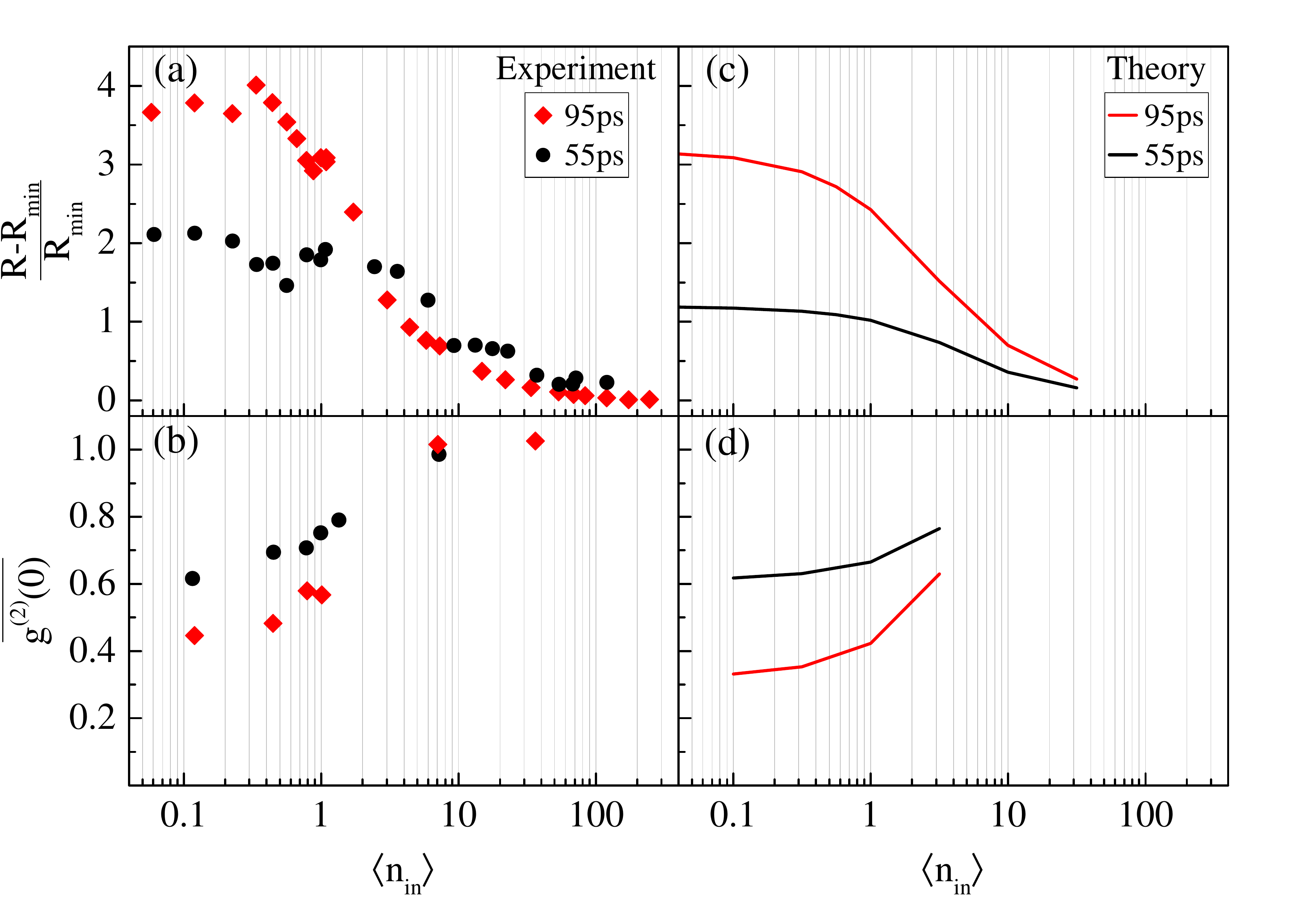}
\end{center}
\caption{{\bf Effects of the temporal length of the excitation pulses on the Nonlinearity curve and Second-order correlation measurements as function of $\langle n_{\textrm{in}}\rangle$.} (a) Measured nonlinearity for $55$ ps (black circles) and $95$ ps (red diamonds) temporal length of the excitation pulse. (b) Measured second-order correlation measurements at zero delay for $55$ ps (black circles) and $95$ ps (red diamonds) temporal length of the excitation pulse. (c),(d) Corresponding calculated curves of the nonlinearity/$\overline{g^{(2)}(0)}$ for $55$ ps (black line) and $95$ ps (red line) temporal length of the excitation pulse.}
\label{fig:figpulses}
\end{figure}

We present here experimental data measured on another QD-cavity device presenting the following parameters: $\eta_{\textrm{top}}{=}63.5\%$,  $g{=}19$ $\mu$eV \ , $\gamma{=}0.5 \mu$eV and  $\kappa{=}100 \mu$eV . The  exciton fine structure splitting for this QD is $\Delta_{\mathrm{FSS}}{=}10 \mu$eV 
and the relative orientation of the QD and cavity axes $\theta{=}20^\circ$. 
The  cooperativity is $C{=}g^2/(\kappa\gamma) {\approx} 7$. 
Figure \ref{fig:figpulses}(a,b) present the measured and calculated reflectivity contrast $\frac{R-R_{min}}{R_{min}}$ as a function of the average incident photon number  $\langle n_{\textrm{in}}\rangle$ for two different pulse lengths. $R_{min}$  is the reflectivity at saturation.  Fig. \ref{fig:figpulses}(c,d) present the corresponding measured and calculated $\overline{g^{(2)}(0)}$. 

Figure \ref{fig:figpulses}(a) shows how a longer excitation pulse -- and closer to the exciton lifetime -- improves the nonlinear response of the device. Going from a 55 ps to a 95 ps pulse length, the contrast of the curve is roughly doubled and, at the same time, the nonlinearity threshold decreases from $\langle n_{\textrm{in}}\rangle {\approx} 2$ to $\langle n_{\textrm{in}}\rangle {\approx} 0.5$. The narrower spectrum of the 95 ps pulse allows to limit the light directly reflected by the cavity, while exciting more efficiently the QD transition. The same effect is revealed in Fig. \ref{fig:figpulses}(b), where the $\overline{g^{(2)}(0)}$ for the light reflected at low incident photon number achieves a lower value of $0.44$ for the longer pulse, thanks to the presence of a higher fraction of light re-emitted by the QD.

While not shown here, it is important to mention that excessively long pulses have the opposite effects on the $\overline{g^{(2)}(0)}$.
When the pulse length exceeds the exciton lifetime, the probability of  multiple photon emission within the same  pulse increases, degrading the value of the measured $\overline{g^{(2)}(0)}$.

\medskip
\medskip

\noindent {\bf Raw histograms on the three-photon correlation measurements}

\medskip

\begin{figure}[h]
\setlength{\abovecaptionskip}{-5pt}
\setlength{\belowcaptionskip}{-2pt}
\begin{center}
\includegraphics[width=.6\linewidth,angle=0]{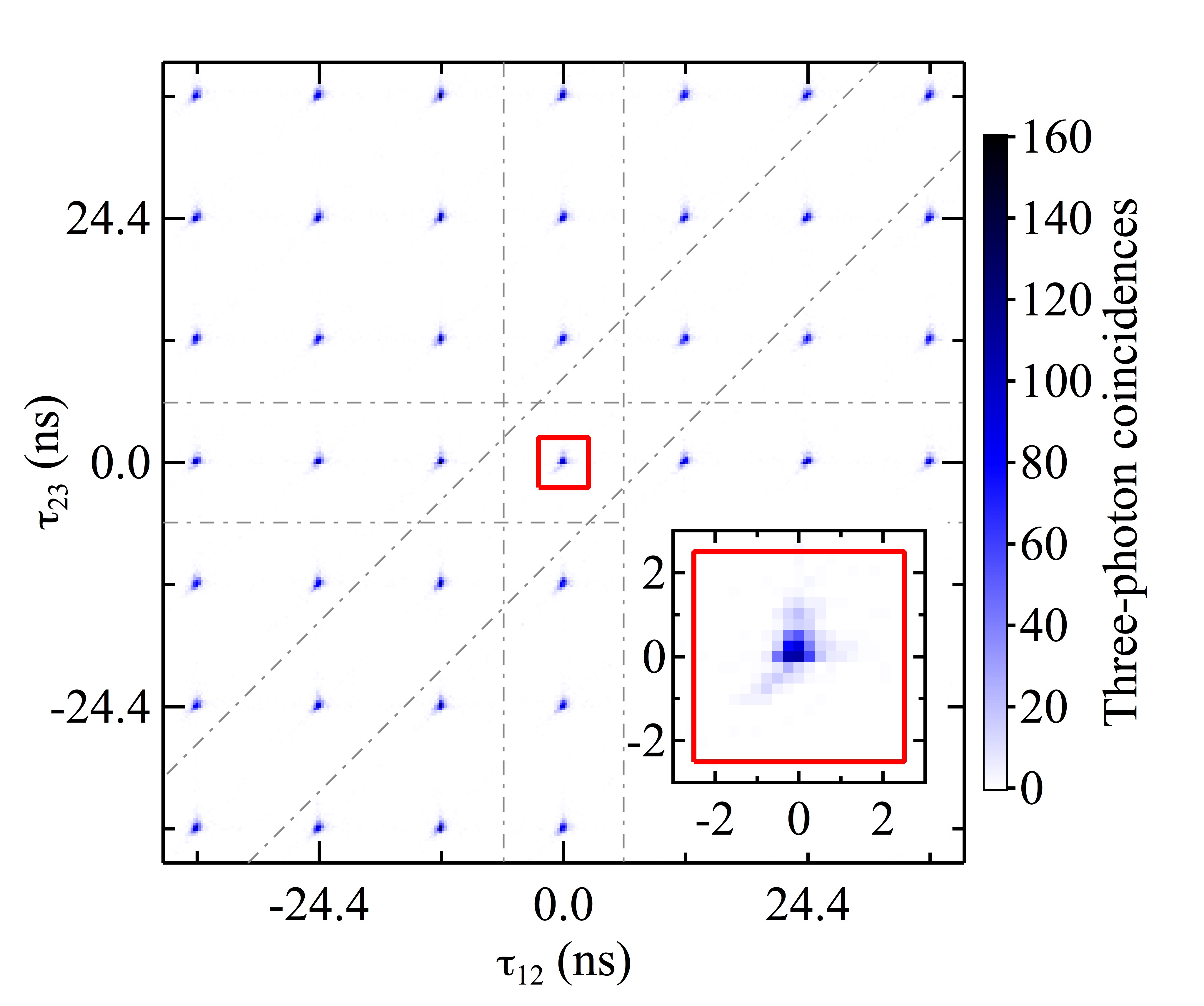}
\end{center}
\caption{Three-photon coincidence map as function of the relative delays between detectors 1-2 (bottom axis) and 2-3 (vertical axis). The two-dimensional time bin for the coincidences detection is $256\times256$ ps$^{2}$.}
\label{fig:figraw}
\end{figure}

Figure \ref{fig:figraw} shows the raw two-dimensional histogram map of Fig. \ref{fig:fig4}(a); it corresponds to the three-photon coincidences of the incident laser (rendering a homogeneous distribution of correlation peaks) detected in SPADs 1, 2 and 3, at different relative delays, $\tau_{12}$ and $\tau_{23}$, accordingly to the description of the detection setup described in Fig. \ref{fig:fig2}(b), configuration C. 

The figure shows the region of the three-photon coincidences up to a maximum relative delay of $\pm3$ pulses, but the measured histograms maps are generated over relative time delays of $\pm0.5$ $\mu$s. 

The integration area for each peak, extended over $5\times5$ ns$^2$, is marked, as an example, with a red line around the zero delay peak. It must be mentioned that the integration area is more than 5 times bigger than the temporal area of the correlation peaks. As described in the Methods section, we work in the Time-Tagged Time-Resolved mode of the  correlator, and we choose a three-fold coincidences time bin of 256 ps, clearly visible in the bottom-right inset of the figure. The asymmetry of the peaks arises from the slightly different response time of each SPAD.

\medskip
\medskip

\noindent {\bf Theory}

\medskip

\noindent The QD is modeled as a three-level system involving the ground state \tket{G} and two linear excitons \tket{X} and \tket{Y} of respective transition frequencies $\omega_X$ and $\omega_Y$ and fine-structure splitting $\Delta_\mathrm{FSS}$. The spontaneous emission and dephasing rates 
(taken equal for both excitons) are respectively denoted $\gamma_{sp}$, and $\gamma^*$. 
We have considered two quasi-resonant modes of the cavity $a_H$ and $a_V$ of respective polarizations $H$ and $V$. Both modes have the same width $\kappa$ and are equally coupled to both excitonic transitions with the parameter $g$. The angle between the QD's  and cavity's natural axes is denoted $\theta$. Finite spatial overlap between the cavity and the driving field is taken into account through the parameter $\eta_{in}$, while the mirrors' finite transmittances and losses are modeled through the parameter $\eta_{top}$. 
The $H$-polarized driving light of frequency $\omega$ is modeled by a classical, time-dependent Hamiltonian.

We have solved the Lindblad master equation for the density matrix $\rho$ of the full QD-cavity system:
\begin{figure}
\setlength{\abovecaptionskip}{-5pt}
\setlength{\belowcaptionskip}{-2pt}
\begin{center}
\includegraphics[width=.45\linewidth,angle=0]{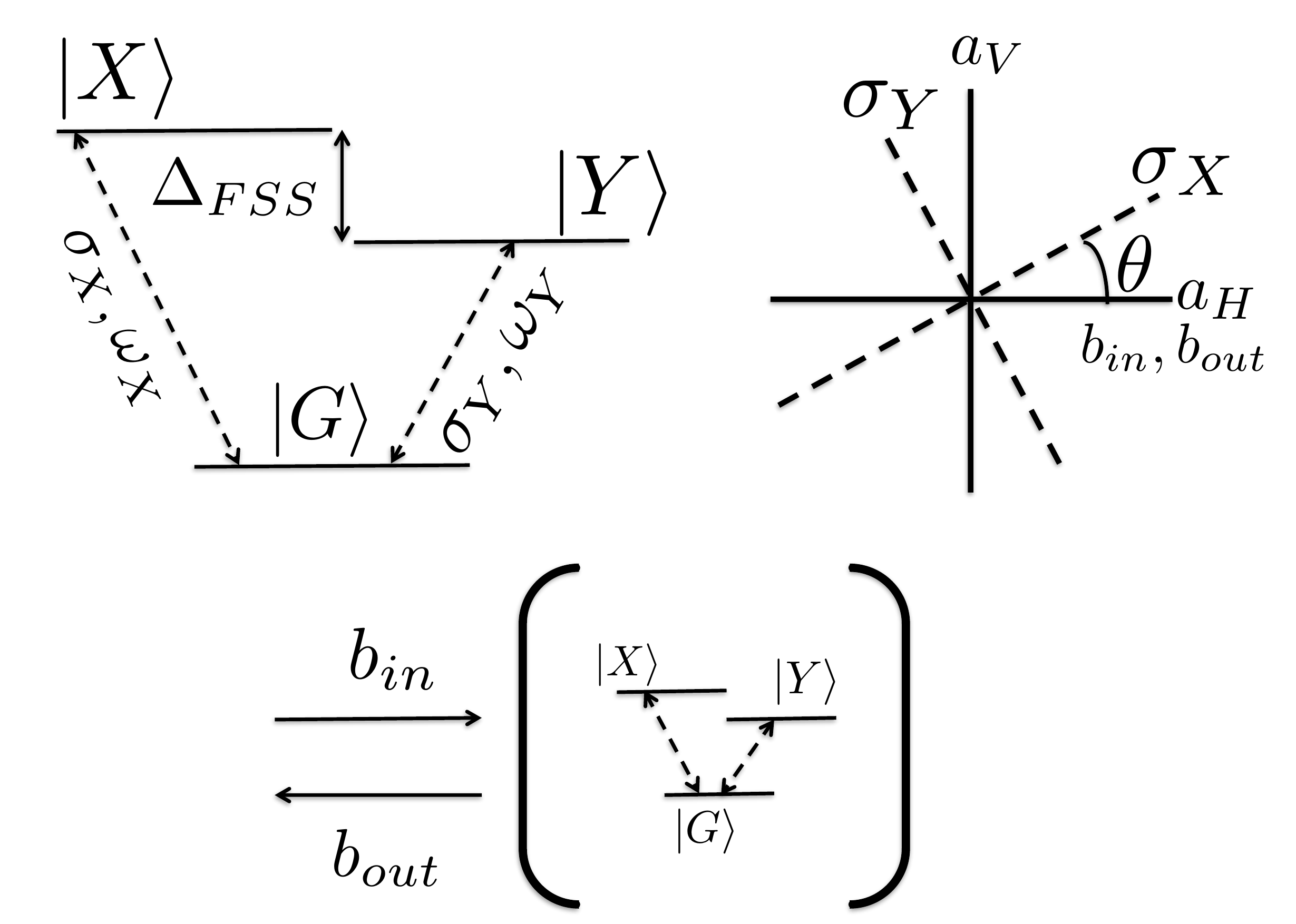}
\end{center}
\caption{ Supplementary figure 2: schematic of the model and notations. (a)  Three-level atomic system. (b) Relative orientations of the QD exciton axes and the cavity axes. (c) Input, output and cavity field operators.}
\label{figmodel}
\end{figure}

\begin{equation} \label{L}
\dot{\rho}=\mathcal{L}[\rho]=-\frac{i}{\hbar}\left[\hat{H},\rho\right]+ D_{\gamma_{sp},\sigma_H}[\rho] + D_{\gamma^*,\Pi_{ex}}[\rho] + D_{\gamma_{sp},\sigma_V}[\rho] + D_{\kappa,a_H}[\rho] + D_{\kappa,a_V}[\rho] 
\end{equation}
\noindent $D_{\alpha,X}[\rho] = \alpha \left( X \rho X^\dagger - \frac{1}{2} (X^\dagger X \rho + \rho  X^\dagger X) \right)$
is the Lindbladian super-operator describing the relaxation or the pure dephasing, involving the QD's operators $\sigma_V = \ket{G}\bra{V}$ and $\sigma_H=\ket{G}\bra{H}$. We have introduced the respective Hamiltonian of the problem

\begin{align} \label{H}
\hat{H}=\hat{H}_\text{QD}+\hat{H}_\text{c}+\hat{H}_\text{i}+\hat{H}_\text{p} \\
\hat{H}_\text{QD}=\hbar (\delta_V^\mathrm{QD} \sigma_V^\dag\sigma_V+\delta_H^\mathrm{QD} \sigma_H^\dag\sigma_H)- \Delta_\mathrm{FSS} \cos(\theta) \sin(\theta) (\sigma_H^\dag\sigma_V+\sigma_V^\dag\sigma_H) \\
\hat{H}_\text{c}=\hbar( \delta_V a_V^\dag a_V+\delta_H a_H^\dag a_H )\\
\hat{H}_\text{i}=-i \hbar g(a_V \sigma_V^\dag+a_H \sigma_H^\dag-a_V^\dag \sigma_V-a_H^\dag \sigma_H) \\
\hat{H}_\text{p}(t)=i \hbar (\Omega^*(t) a_H - \Omega(t) a_H^\dag)
\end{align}
\noindent $\hat{H}_\text{QD}$ is the free Hamiltonian of the QD, written as a function of the QD states

\begin{align}
\ket{V} = \cos(\theta) \ket{X} + \sin(\theta) \ket{Y} \\
\ket{H} = -\sin(\theta) \ket{X} + \cos(\theta) \ket{V}
\end{align}

\noindent of respective energies $\delta_H^\mathrm{QD}=\delta_X \sin^2(\theta)+\delta_Y\cos^2(\theta)$ and $\delta_V^\mathrm{QD}=\delta_X \cos^2(\theta)+\delta_Y \sin^2(\theta)$. 
$\delta_X = \omega_X - \omega$ and $\delta_Y=\omega_Y-\omega$ are the respective detunings of each excitonic transition w.r. to the pump frequency, and $\Pi_{ex}=\ket{H}\bra{H}+\ket{V}\bra{V}$. $\hat{H}_\text{c}$ is the free Hamiltonian of the cavity modes.
$ \hat{H}_\text{i}$ is QD-cavity interaction.

The classical drive is induced by some $H$-polarized field injected in the input port of the cavity mode. It acts on the QD through the Hamiltonian $\hat{H}_\text{p}(t)$ with classical Rabi frequency
\begin{align}
\Omega(t)=\sqrt{\eta_{top}\kappa} \langle \hat{b}_{in} \rangle (t) \\
\langle b_{in}(t) \rangle=\sqrt{n_{in}} \left(\dfrac{4 \ln(2)}{\pi \tau^{2}}\right)^{1/4}\exp(-2\ln(2){t^2}/{\tau^2})
\end{align}
\noindent where $\hat{b}_{in}$ is the input field operator, $\langle a \rangle=\text{Tr}\left(a \rho\right)$ for any operator $a$, and $n_{in}$ is the mean number of photons in the pulse.

Finally, the $H$-polarized detected field operator $\hat{b}_{out}$ verifies the standard input-output equation:
\begin{align}\label{out_norm}
\hat{b}_{out}=\hat{b}_{in}+\sqrt{\eta_{top}\kappa}\,a_H
\end{align}

The theoretical value of $\overline{g^{(2)}(0)}$ was computed using the following formula:
\begin{align}\label{g2}
\overline{g^{(2)}(0)}=\frac{\int^\infty_{-\infty} dt \int^\infty_{-\infty} d\tau G^{(2)}(t,t+\tau)}{[\int^\infty_{-\infty} dt \langle \hat{b}_{out}^\dag(t) \hat{b}_{out}(t)\rangle]^2}
\end{align}
where 
\begin{align}
G^{(2)}(t_1,t_2)=\langle \hat{b}_{out}^\dag(t_1)\hat{b}_{out}^\dag(t_2)\hat{b}_{out}(t_2)\hat{b}_{out}(t_1)\rangle
\end{align}
The two-time correlations functions are derived using the Quantum Regression Theorem, such that $\langle a^\dag(t) a^\dag(t+\tau) a(t+\tau) a(t)\rangle=\text{Tr}\left[ \left(U(t,t+\tau) a \rho(t) a^\dag U^\dag(t,t+\tau)\right)  a^\dag a\right]$. We have introduced the 
evolution super-operator $U(t_1,t_2)$, verifying  $\rho(t_2)=U(t_1,t_2) \rho(t_1) U^\dag(t_1,t_2)=\int_{t_1}^{t_2} d t^\prime \mathcal{L}[\rho(t^\prime)]$, where the expression of $\mathcal{L}[\rho(t^\prime)]$ is given in (\ref{L}).\\


\end{document}